\documentclass[10pt,english]{iopart}
\usepackage[T1]{fontenc}
\usepackage[latin1]{inputenc}
\usepackage{fancyhdr}
\usepackage{amstext}
\usepackage{graphicx}
\usepackage[numbers]{natbib}

\makeatletter

\newcommand{\noun}[1]{\textsc{#1}}
\providecommand{\tabularnewline}{\\}

\usepackage{iopams}
\usepackage{setstack}

\usepackage{subfigure}

\makeatother

\usepackage{babel}
\begin{document}

\title{Structure identification methods for atomistic simulations of crystalline
materials}

\author{Alexander Stukowski}

\address{Lawrence Livermore National Laboratory, Livermore, CA 94550, USA}

\ead{alexander@stukowski.de}
\begin{abstract}
We discuss existing and new computational analysis techniques to classify
local atomic arrangements in large-scale atomistic computer simulations
of crystalline solids. This article includes a performance comparison
of typical analysis algorithms such as \emph{Common Neighbor Analysis},
\emph{Centrosymmetry Analysis}, \emph{Bond Angle Analysis}, \emph{Bond
Order Analysis}, and \emph{Voronoi Analysis}. In addition we propose
a simple extension to the \emph{Common Neighbor Analysis} method that
makes it suitable for multi-phase systems. Finally, we introduce a
new structure identification algorithm, the \emph{Neighbor Distance
Analysis}, that is designed to identify atomic structure units in
grain boundaries.

\noindent \begin{flushleft}
\vspace{1.5cm}
\textsf{This paper has been published in }\textsf{\emph{Modelling
and Simulation in Materials Science and Engineering}}\textsf{ 20 (2012),
045021.}
\par\end{flushleft}
\end{abstract}
\maketitle

\section{Introduction}

Atomistic simulation methods such as molecular dynamics (MD), molecular
statics, and Monte Carlo schemes are routinely used to study crystalline
materials at the atomic scale. In many cases crystal defects play
a critical role in materials behavior, and their identification in
the simulation data is essential for the understanding of materials
properties. Classical atomistic simulation models, however, do not
keep track of crystal defects explicitly. These models are governed
by a Hamiltonian or other rules which determine the trajectories of
individual particles. Therefore, crystal defects and defect-free crystal
regions must be recovered from the generated particle-position datasets
in a post-processing step to enable the interpretation of simulation
results.

For this purpose, many computational analysis methods have been developed
in the past. Their task is to assign a structural type to each atom
or particle based on an analysis of its local environment. Most such
methods attempt to match a local structure to an idealized one (such
as fcc or bcc), and measure how closely they fit. This information
can then be used to color particles for visualization purposes or
to quantify the occurrence of different crystalline phases and defects
in a simulation. Another important application is filtering the simulation
data on the fly to reduce it to a manageable amount, e.g. by storing
only particles with an atypical environment.

Our goal is to give an overview of current computational analysis
techniques, as they are offered by many visualization tools and simulation
codes, and as they are employed in many recent simulation studies
described in the literature. In particular we will review the most
commonly used structure characterization methods for molecular dynamics
simulations of crystalline solids:
\begin{itemize}
\item (a) energy filtering, 
\item (b) centrosymmetry parameter analysis (CSP) \citep{Kelchner1998}, 
\item (c) bond order analysis \citep{Steinhardt1983}, 
\item (d) common neighbor analysis (CNA) \citep{Honeycutt1987}, 
\item (e) bond angle analysis (BAA) \citep{Ackland2006}, and
\item (f) Voronoi analysis. 
\end{itemize}
In addition to describing the respective strengths and weaknesses
of these methods, we introduce two new identification schemes:
\begin{itemize}
\item (g) adaptive common neighbor analysis (a-CNA) and
\item (h) neighbor distance analysis (NDA).
\end{itemize}
The adaptive CNA is a simple extension of the common neighbor analysis
method to improve the characterization of multi-phase systems. The
computationally more expensive NDA is targeted at the classification
of complex structural environments as they occur inside crystal defects
such as grain boundaries. 

As part of this paper we have implemented all discussed algorithms
for benchmarking purposes. We provide the source code of this analysis
tool at http://asa.ovito.org/ as a reference, to facilitate the use
of the described techniques, and to foster their advancement by the
research community.

\section{Existing analysis methods\label{sec:ExistingTechniques}}

Here, we focus on analysis techniques for simulation studies of crystalline
solids only. A broader overview of structural characterization methods
and shape matching algorithms for general particle systems has recently
been given by Keys et al. \citep{Keys2011}.

\subsection{General considerations}

One can name several features that an ideal structure characterization
technique should provide: 
\begin{itemize}
\item Accuracy - The method should be able to correctly distinguish several
structural environments solely based on the local arrangement of atoms
and independent of the crystal orientation (rotational and translational
invariance).
\item Robustness - The algorithm should assign a local structure to most
particles in the system and avoid errors arising from small displacements
of particles from their equilibrium or symmetry positions.
\item Computational efficiency - Since the local structure characterization
needs to be performed for every particle in a system, and possibly
at high frequency as part of an on-the-fly analysis, the computational
cost is an important factor.
\item Simplicity - Because wide-spread use of a method requires an algorithm
that is easy to implement and understand.
\item Universality - Ideally, the set of reference structures recognized
by the method is not hard-coded into the algorithm and can be easily
extended by the user.
\end{itemize}
Note that the first two requirements are in conflict with each other:
A low sensitivity to atomic displacements usually comes at the price
of a reduced capability of the identification method to distinguish
similar structures. Some of the methods discussed here allow the user
to explicitly control this tradeoff between accuracy and robustness.
In general one wants to avoid any wrong classifications, i.e. \emph{false
positives} as well as \emph{false negatives}, in the structure recognition
process.

The techniques discussed in this paper can be divided into two sets.
Methods (a)-(c) quantify the similarity of a given atomic arrangement
to a particular reference structure. A positive match is detected
by comparing the computed similarity measure to a threshold chosen
by the user. A high threshold increases the robustness (and the chance
of false positives) while a low threshold increases the sensitivity
(and the chance of false negatives). The aim of the second group of
methods is to distinguish between several reference structures and
to uniquely assign a type to each particle in the system (with the
possibility of assigning no type at all if the local atomic arrangement
deviates too much from all of the reference structures). These structure
identification methods are usually based on a discrete signature that
is calculated from the particle positions, and which identifies the
arrangement unambiguously.

\subsection{Energy filtering}

The potential energy of an atom can be used as a simple indicator
to decide whether it forms a perfect lattice with its neighbors. Given
that atoms which are part of a crystal defect are usually higher in
energy than the perfect lattice (the ground state), one can detect
defective atoms by using a simple threshold criterion: Atoms having
a potential energy above the threshold are considered defect atoms,
while low-energy atoms are classified as regular crystalline atoms.

Several shortcomings of this method have contributed to the fact that
it is rarely used nowadays. The atomic energy levels of perfect lattice
atoms and metastable defects can easily overlap due to degeneracies,
elastic strain energy, or thermal energy. Then the discrimination
between the different structural states becomes impossible. Moreover,
the potential energy of individual atoms is specific to the employed
interaction model, and, for quantum mechanical descriptions and some
interatomic potentials, is not defined at all. This is why one prefers
purely structural analysis methods, which characterize the spatial
arrangement of atoms without reference to the interatomic interaction
laws.

\subsection{Centrosymmetry parameter}

The centrosymmetry property of some lattices (e.g. fcc and bcc) can
be used to distinguish them from other structures such as crystal
defects where the local bond symmetry is broken. Kelchner et al. \citep{Kelchner1998}
have developed a metric, the so-called centrosymmetry parameter\emph{
}(CSP), that quantifies the local loss of centrosymmetry at an atomic
site, which is characteristic for most crystal defects. The CSP of
an atom having $N$ nearest neighbors is defined as

\begin{equation}
\textrm{CSP}=\sum_{i=1}^{N/2}\left|\mathbf{r}_{i}+\mathbf{r}_{i+N/2}\right|^{2}\label{eq:csp}
\end{equation}
where $\mathbf{r}_{i}$ and $\mathbf{r}_{i+N/2}$ are vectors from
the central atom to a pair of opposite neighbors. Practical ways of
finding these pairs are described in \citep{BulatovCaiComputerSimDisloc}
and in the accompanying documentation of the visualization program
\noun{AtomEye} \citep{Li2003AtomEye} and the molecular dynamics code
\noun{LAMMPS} \citep{Plimpton1995}. The latter uses the following
scheme: There are $N(N-1)/2$ possible neighbor pairs $(i,j$) that
can contribute to above formula. The sum of two bond vectors, $\left|\mathbf{r}_{i}+\mathbf{r}_{j}\right|^{2}$,
is computed for each, and only the $N/2$ smallest are actually used
to compute the CSP. For centrosymmetric lattice sites, they will be
pairs of neighbor atoms in symmetrically opposite positions with respect
to the central atom; hence the $i+N/2$ notation in formula \ref{eq:csp}.
The CSP is close to zero for regular sites of a centrosymmetric crystal
and becomes non-zero for defect atoms. The number of nearest neighbors
taken into account is $N=12$ for fcc and $N=8$ for bcc.

The main advantage of the CSP is that it is only marginally affected
by elastic distortions of the crystal. In particular, any affine deformation
of the lattice does not change its degree of centrosymmetry at all.
The CSP is, however, sensitive to random thermal displacements of
atoms. Being only a scalar measure, the CSP's capability of discriminating
between different defect types is rather weak. The noise induced by
thermal displacements and inhomogeneous elastic strain may well dominate
any characteristic differences between defect structures. Most notably,
the method can only be applied to the class of centrosymmetric lattices
(which does not include hcp, for example), and it provides no means
of distinguishing multiple centrosymmetric crystal phases.

\begin{figure}
\centering{}\includegraphics[width=0.5\textwidth]{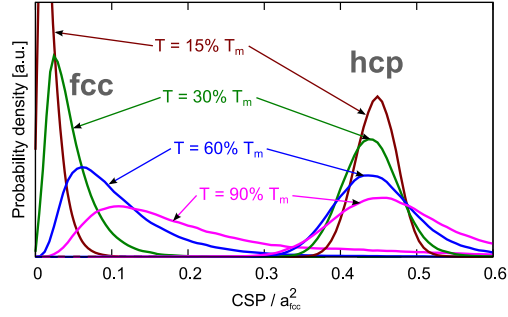}\caption{Distribution of CSP values (normalized by the square of the lattice
parameter) measured at various homologous temperatures (in a Cu crystal).
The CSP values have been sampled from atoms located in the defect-free
fcc lattice and in a fcc stacking fault (corresponding to an hcp-like
arrangement of neighbors with broken centrosymmetry).}
\label{fig:csp_distribution}
\end{figure}

The user needs to choose a proper threshold to distinguish defect
atoms from perfect lattice atoms. At elevated temperatures the distribution
of CSP values in a perfect crystal becomes broader, and may begin
to overlap with the characteristic range of crystal defects. To show
this, we have measured the CSP distributions in a perfect fcc Cu crystal
and inside an intrinsic stacking fault at various temperatures (Fig.
\ref{fig:csp_distribution}). While at low homologous temperature
the two distributions are well-separated, the identification of perfect
fcc atoms becomes less reliable at high temperature (above 60\% of
the melting temperature). Note that other crystal defects such as
Shockley partial dislocations exhibit CSP values lower than those
of stacking fault atoms.

\subsection{Bond order analysis}

Given a central atom, we can project its near neighbor bonds to a
unit sphere. Based on these projected vectors one can define a set
of local bond order parameters, also known as \emph{Steinhardt order
parameters} \citep{Steinhardt1983}, that are rotationally invariant
combinations of spherical harmonics. The bond order parameters exhibit
characteristic values for each crystal structure, allowing us to discriminate
between them.

Given the $N$ neighbor vectors $\mathbf{r}_{1}\ldots\mathbf{r}_{N}$
of a central atom, the $l$-th order parameter is defined according
to Steinhardt as
\begin{equation}
Q_{l}=\sqrt{\frac{4\pi}{2l+1}\sum_{m=-l}^{+l}\left|q_{lm}\right|^{2}}
\end{equation}

with 

\begin{equation}
q_{lm}=\frac{1}{N}\sum_{i=1}^{N}Y_{l}^{m}(\mathbf{r}_{i}).
\end{equation}
Here, the complex functions $Y_{l}^{m}(\mathbf{r})=Y_{l}^{m}(\theta,\varphi)$
are the spherical harmonics, whose evaluation is computationally expensive.
Note that the set of local bond order parameters, $\left\{ Q_{l}\right\} $
with $l=1,2,3,\ldots$, is invariant under rotations of the coordinate
system (meaning that it is independent of the crystal's orientation).
Bond order parameters up to $l=3$ are zero for lattices with cubic
symmetry, and one usually takes into account the values of $Q_{4}$
and $Q_{6}$ to discriminate between fcc, hcp and bcc phases. %
\footnote{One obtains the following reference values for the perfect lattices:
$Q_{4}^{\textrm{fcc}}=0.191$, $Q_{6}^{\textrm{fcc}}=0.575$ and $Q_{4}^{\textrm{hcp}}=0.097$,
$Q_{6}^{\textrm{hcp}}=0.485$ and $Q_{4}^{\textrm{bcc}}=0.036$, $Q_{6}^{\textrm{bcc}}=0.511$.%
}

The parameter set $(Q_{4},Q_{6})$ can be used to measure the structural
order of a particle system when averaged over all atoms. Hence, bond
order parameters are often used in computational studies of crystallization
to determine the fractions of crystalline and liquid phases. Crystal
deformation and thermal fluctuations, however, smear out the order
parameter distributions \citep{Lechner2008}. Thus, to assign a particular
structure type to a particle, one needs to define non-overlapping
regions in the $Q_{4}$-$Q_{6}$ parameter plane \citep{Desgranges2008}
for all crystal phases considered. The choice of these regions is
arbitrary, and, to our knowledge, no generally accepted scheme for
the classification of bond order parameters exists so far.

\subsection{Common neighbor analysis\label{sub:Common-neighbor-analysis}}

Structure analysis methods that employ more complex, high-dimensional
signatures to characterize arrangements of atoms are usually better
in discriminating between several structures. A popular method of
this type is the common neighbor analysis\emph{ }(CNA) \citep{Honeycutt1987,Faken1994}.
Unlike the CSP and the local bond order parameters, the CNA does not
directly take into account the spatial vectors pointing from the central
atom to its neighbor. Instead, a characteristic signature is computed
from the topology of bonds that connect the surrounding neighbor atoms.

Usually, two atoms are said to be (near-)neighbors, or bonded, if
they are within a specified cutoff distance $r_{\textrm{cut}}$ of
each other. For densely packed structures (fcc and hcp) the cutoff
distance is set to be halfway between the first and second neighbor
shell, giving for fcc

\begin{equation}
r_{\textrm{cut}}^{\textrm{fcc}}=\frac{1}{2}\left(\sqrt{1/2}+1\right)a_{\textrm{fcc}}\simeq0.854\, a_{\textrm{fcc}},\label{eq:cna_cutoff_fcc}
\end{equation}

where $a_{\textrm{fcc}}$ is the lattice constant of the crystal structure.
For the bcc lattice, two neighbor shells need to be taken into account,
and atoms are considered to be bonded with their first- and second-nearest
neighbors:

\begin{equation}
r_{\textrm{cut}}^{\textrm{bcc}}=\frac{1}{2}\left(1+\sqrt{2}\right)a_{\textrm{bcc}}\simeq1.207\, a_{\textrm{bcc}}.\label{eq:cna_cutoff_bcc}
\end{equation}

To assign a local crystal structure to an atom, three characteristic
numbers are computed for each of the $N$ neighbor bonds of the central
atom: The number of neighbor atoms the central atom and its bonded
neighbor have in common, $n_{cn}$; the total number of bonds between
these common neighbors, $n_{b}$; and the number of bonds in the longest
chain of bonds connecting the common neighbors, $n_{lcb}$. This yields
$N$ triplets $(n_{cn},n_{b},n_{lcb})$, which are compared to a set
of reference signatures to assign a structural type to the central
atom (Table~\ref{tab:CNA}). \medskip{}

\begin{table}[h]
\centering{}%
\begin{tabular}{|c|c|c|c|}
\hline 
fcc ($N=12$) & hcp ($N=12$) & bcc ($N=14$) & cubic diamond ($N=16$)\tabularnewline
\hline 
12 $\times$ (421) & 6 $\times$ (421) & 8 $\times$ (666) & 12 $\times$ (543)\tabularnewline
 & 6 $\times$ (422) & 6 $\times$ (444) & 4 $\times$ (663)\tabularnewline
\hline 
\end{tabular}\caption{CNA signatures of common crystal structures. For example: An hcp-coordinated
atom has six bonds of (421) type and six of (422) type. That is, any
two near-neighbors in a hcp crystal have exactly four common neighbors,
which are interconnected by two bonds. And the longest continuous
chain these two bonds form is either of length one or two (in six
cases each).}
\label{tab:CNA}
\end{table}
\medskip{}

The \emph{common neighborhood parameter} \citep{Tsuzuki2007} should
be mentioned as an alternative analysis method, which was proposed
by Tsuzuki et al. to combine the strengths of both the CNA and CSP
methods. The CNA has also been extended to binary atomic systems by
taking the chemical species of common neighbors into account as an
additional criterion \citep{Lummen2007}. This extension enables the
identification of simple binary structures such as $\textrm{L}1_{0}$,
$\textrm{L}1_{2}$ etc.

\subsection{Bond angle analysis}

The bond angle analysis\emph{ }has been developed by Ackland and Jones
\citep{Ackland2006} to distinguish fcc, hcp and bcc coordination
structures. From the $N$ bond vectors of the central atom, an eight-bin
histogram of the $N(N-1)/2$ bond angle cosines, $\cos\theta_{ijk}$,
is computed first. Here, $\theta_{ijk}$ denotes the angle formed
by the central atom $i$, and two of its neighbors, $j$ and $k$.
The obtained histogram is then further evaluated using a set of heuristic
decision rules to determine the most likely structure type. These
rules have been optimized by the authors such that a robust discrimination
of the most important crystal structures is archived. The number of
neighbors used to calculate the bond angle distribution is determined
adaptively by employing a cutoff radius that is proportional to the
average distance of the six nearest neighbors.

\subsection{Voronoi analysis\label{sub:Voronoi}}

The Voronoi decomposition \citep{Voronoi1908} can serve as a geometric
method to determine the near neighbors of a particle (i.e. its coordination
number) by considering the faces of the Voronoi polyhedron enclosing
the particle. Furthermore, the geometric shape of the Voronoi polyhedron
reflects the characteristic arrangement of near neighbors. For this
reason the Voronoi decomposition has been employed in simulation studies
of liquids and glasses to analyze various properties of their atomic
structure \citep{Finney1970,Okabe2000}. 

To effectively characterize the arrangement of near neighbors, the
computed Voronoi polyhedron for a particle is translated into a compact
signature by counting the number of polygonal facets having three,
four, five and six vertices/edges. This yields a vector of four integers,
$(n_{3},n_{4},n_{5},n_{6})$, that identifies the structural type.
For instance the Voronoi polyhedron of an fcc lattice atom is equivalent
to the fcc Wigner-Seitz cell and comprises 12 facets with four vertices
each. Thus the corresponding signature for fcc is (0,12,0,0). The
polyhedron of a bcc atom has facets with four and six vertices, and
the corresponding signature is (0,6,0,8).

Even though the Voronoi method has been used numerous times for the
analysis of particle systems without long-range order such as liquids
and glasses, it has rarely been applied to simulations of crystalline
materials. One reason is that singular Voronoi vertices, which are
adjacent to more than three facets, and which occur in the Voronoi
decomposition of some highly symmetric crystalline packings such as
fcc and hcp, will dissociate into multiple vertices as soon as the
atomic coordinates are only slightly perturbed. This dramatically
changes the Voronoi polyhedra and the computed signatures \citep{Hsu1979},
making the identification of such crystal structures nontrivial.

In our implementation we use the following approach to mitigate the
problem of singularities in the Voronoi decomposition of fcc crystals:
First, the conventional Voronoi polyhedra are constructed (using the
\textsc{\noun{Voro++}} code library \citep{Rycroft2009}). When counting
the number of edges of a Voronoi facet, we skip edges which are shorter
than a certain threshold. Thus, a singular vertex, which may have
dissociated into multiple vertices due to perturbations, will still
be counted as one. Small facets with less than three edges above the
threshold are completely ignored. The edge threshold is set to 30\%
of the polyhedron's maximum radius.

It should be pointed out that the described sensitivity of the Voronoi
method to perturbations of the atomic coordinates, in addition to
the high computational cost of the Voronoi polyhedron construction,
render its application to crystalline systems rather unattractive.
Remarkably, the Voronoi method, in its simplest form described here,
is not capable of discriminating hcp-coordinated atoms from fcc atoms
because the corresponding Voronoi polyhedra have a (0,12,0,0) signature
in both cases.

\section{New analysis methods}

We now describe two new methods, which can provide superior analysis
results for some applications. The \emph{adaptive common neighbor
analysis} is a simple extension of the standard CNA method, which
adds some convenience on the user's side and the ability to analyze
multi-phase systems. The \emph{neighbor distance analysis} (NDA),
in contrast, is a completely new algorithm that employs a more complex
signature to identify a wider range of atomic arrangements.

\subsection{Adaptive common neighbor analysis}

The common neighbor analysis method described in section \ref{sub:Common-neighbor-analysis}
is one of the most frequently used structure identification methods
for atomistic simulation studies of fcc, hcp, and bcc crystal plasticity.
It provides efficient and unambiguous classification of local atomic
arrangements, making it possible to effectively distinguish crystal
defects from undisturbed lattice atoms. The only parameter required
is the cutoff radius, which determines the maximum separation of near-neighbors,
and which must be chosen according to the crystal phase under consideration
(cf. Eqs. \ref{eq:cna_cutoff_fcc} and \ref{eq:cna_cutoff_bcc}).

In the case of multi-phase systems, however, the choice of the cutoff
parameter is no longer well-defined. In many cases, for instance a
fcc-bcc bicrystal simulation, one cannot specify a global cutoff radius
that fits all phases equally well. We therefore propose to pick the
cutoff radius individually for each atom and in dependence of the
reference structure we want to compare it to. We implement this approach,
which we refer to as \emph{adaptive common neighbor analysis} (a-CNA),
as follows.

Given a central atom to be analyzed, we first generate the list of
$N_{\textrm{max}}$ \emph{nearest} neighbors and sort it by distance.
$N_{\textrm{max}}$ is the maximum required number of neighbors for
all considered reference structures, e.g. $N_{\textrm{max}}=16$ for
the set of structures listed in Table \ref{tab:CNA}. One can generate
such a nearest neighbor list either by means of a $k$-$d$ tree data
structure \citep{Bentley1975} and a recursive $k$-th nearest neighbor
query algorithm \citep{Friedman1977}, or by sorting a pre-existing
neighbor list that has already been generated on the basis of an excessively
large cutoff radius (for instance, to compute the interatomic forces
in a molecular dynamics simulation).

To test whether the local coordination structure matches an fcc crystal,
we take only the first $N_{\textrm{fcc}}=12$ entries from the sorted
neighbor list. The average distance of these 12 nearest neighbors
provides a local length scale, analogous to the approach used in the
bond angle analysis. That is, we can define a local cutoff radius,
which is specific to the current atom and used for matching with the
fcc reference structure only:

\begin{equation}
r_{\textrm{cut}}^{\textrm{local}}(\textrm{fcc})=\frac{1+\sqrt{2}}{2}\cdot\frac{\sum_{j=1}^{12}\left|\mathbf{r}_{j}\right|}{12}.
\end{equation}
The local cutoff is subsequently used to determine the {}``bonding''
between the 12 nearest neighbors and to compute the CNA signature
as usual. If the signature does not conform to fcc, the algorithm
proceeds with testing against the next candidate structure. For the
bcc structure, for instance, the 14 nearest neighbors must be taken
into account and a local cutoff is computed as 

\begin{equation}
r_{\textrm{cut}}^{\textrm{local}}(\textrm{bcc})=\frac{1+\sqrt{2}}{2}\left[\frac{2}{\sqrt{3}}\cdot\frac{\sum_{j=1}^{8}\left|\mathbf{r}_{j}\right|}{8}+\frac{\sum_{j=9}^{14}\left|\mathbf{r}_{j}\right|}{6}\right].
\end{equation}
Here, the local length scale is determined from the eight nearest
neighbors in the sorted neighbor list (forming the first shell) and
the successive six neighbors (forming the second shell). Their average
distances are weighted accordingly to yield the local cutoff radius
that lies halfway between the second and third bcc coordination shell.

\begin{figure}[h]
\begin{centering}
\includegraphics[width=1\textwidth]{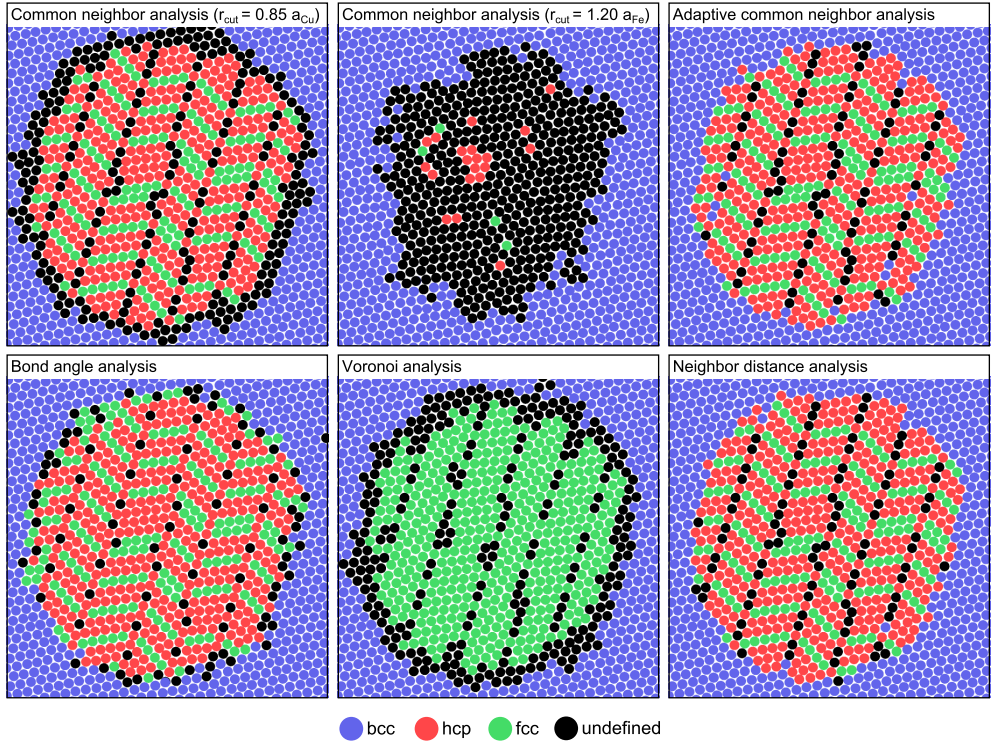}
\par\end{centering}

\caption{Cross-sectional view of a Cu-rich 9R precipitate in bcc-Fe (viewing
direction $\left\langle 111\right\rangle _{\textrm{bcc}}$). The same
simulation snapshot has been analyzed with several coordination structure
identification methods discussed in this paper. Atom colors indicate
the local structural type as classified by the algorithms.}
\label{fig:9R_particle}
\end{figure}

The computational cost of the adaptive CNA increases with the number
of reference structures to be tested. In practice, however, the analysis
is only 25\% more expensive than the standard CNA when identifying
fcc, hcp, and bcc atoms. This cost is in most cases offset by the
convenience of a parameterless method (no cutoff radius) and the superior
analysis results provided. To demonstrate the strength of the adaptive
CNA we have applied it to a simulation of the Fe-Cu multi-phase alloy.
In the simulation study, a combination of Monte Carlo sampling (variance-constrained
semi-grandcanonical ensemble \citep{SadighErhartEtAl2010}) and conventional
MD time integration was used to determine the equilibrium structure
of Cu-rich precipitates in a Fe-rich bcc matrix. The equilibrium distribution
of Cu atoms at a prescribed temperature is found via Monte Carlo transmutation
steps, while alternating MD steps allow the positional degrees of
freedom to relax simultaneously. This enables structural phase transformations
to occur in the simulation. Starting off from a random distribution
of Cu atoms in the bcc-Fe matrix, the Cu atoms precipitate to form
a spherical particle. At certain conditions, the crystal structure
of the cluster changes from bcc to a multiply-twinned 9R structure
(herringbone structure) \citep{Ernst1992} as shown in figure \ref{fig:9R_particle}.
The system has been quenched to zero temperature to remove thermal
displacements prior to the structure analysis.

The results of the conventional common neighbor analysis strongly
depend on the cutoff parameter used. A cutoff that is suitable for
identifying the 9R phase is given by Eq. \ref{eq:cna_cutoff_fcc}
and the lattice constant of fcc-Cu, while for the identification of
the bcc-Fe phase one would apply formula \ref{eq:cna_cutoff_bcc}.
Varying the cutoff between these limiting cases lets the observed
bcc-9R interface slide and makes the precipitate appear smaller or
larger. In all cases, the CNA will be unable to assign a structural
type to the atoms right at the interface since their coordination
does not match to either of the reference structures. The adaptive
CNA overcomes this problem by computing a cutoff radius for each atom
and taking into account the local dilatation. Virtually every atom
in the bcc-9R interface is identified as crystalline by the a-CNA,
giving even slightly better analysis results than the bond angle analysis,
which was specifically designed for applications like this.

\subsection{Neighbor distance analysis\label{sub:Neighbor-distance-analysis}}

In section~\ref{sec:ExistingTechniques} we described several structure
matching methods that all exploit structural symmetries in some way.
Instead of directly comparing the actual atomic positions to a set
of reference coordinates, they condense the particle coordinates into
characteristic signatures which are invariant under rotation, and
which can easily be compared. This transformation is essentially what
makes the identification process efficient and robust (see~\citep{Keys2011}
for an in-depth discussion). Note that, at the same time, this data
reduction usually results in some insensitivity to elastic deformations:
Small perturbations of the atomic positions do not change the calculated
signature.

In general, however, the atomic arrangements found in the core regions
of crystal defects such as grain boundaries may not exhibit any symmetries
or order (e.g. discrete neighbor shells). It might therefore be more
difficult to reduce their description to a small, but unambiguous
set of characteristic numbers (and even less so to a scalar signature
like the CSP). Thus, in such a case, one has to resort to more extensive
types of signatures, as we will propose it in the following. Here,
we will introduce the \emph{neighbor distance analysis} (NDA), a new
structure identification method that aims at situations where the
coordination structure of atoms is lacking any particular symmetries
or shell structure that could be exploited, as it is often the case
in crystal defect cores. 

Let us assume that a reference coordination structure (which we want
to search for in the simulation data) is specified in terms of the
list of bond vectors $(\mathbf{R}_{1},\ldots,\mathbf{R}_{N})$ connecting
the central atom with its $N$ nearest neighbors (with $N$ being
a freely selectable parameter). The coordination pattern for fcc lattice
atoms, for instance, would consist of $N=12$ neighbors, with the
reference vectors $(\mathbf{R}_{1},\ldots,\mathbf{R}_{N})$ comprising
the $1/2\left\langle 110\right\rangle $ vector family. We assume
that this list of vectors is ordered according to their distance from
the central atom such that $R_{1}\leq\ldots\leq R_{N}$. 

Given an atom to be analyzed and to be tested against the reference
pattern described above, we first determine its $N$ nearest neighbor
vectors, $(\mathbf{r}_{1},\ldots,\mathbf{r}_{N})$, and sort them
according to their magnitude (such that $r_{1}\leq\ldots\leq r_{N}$).
Obviously, this is not sufficient to associate the actual neighbor
bonds with their counterparts in the reference pattern: The bond lengths
may be perturbed by thermal displacements, and the ordering can be
non-unique if neighbors are arranged on shells. Despite that, we may
compute a local hydrostatic scale factor, $\lambda$, from the two
sorted bond lists:
\begin{equation}
\lambda=\frac{1}{N}\sum_{i=1}^{N}\left(R_{i}/r_{i}\right).\label{eq:scaling_factor}
\end{equation}
This scale factor relates the lattice constant of the reference structure
(which is arbitrary, and may be chosen to be unity) to that of the
actual crystal, which depends on factors such hydrostatic stress,
temperature, and chemical composition.

The one-to-one mapping between the reference vectors $(\mathbf{R}_{1},\ldots,\mathbf{R}_{N})$
and the actual neighbor vectors $(\mathbf{r}_{1},\ldots,\mathbf{r}_{N})$,
as we still need to determine it, can be expressed in terms of a permutation
$\sigma=(\mathbf{r}_{\sigma(1)},\ldots,\mathbf{r}_{\sigma(N)})$ of
the original neighbor list. Note that multiple equivalent permutations
may exist due to symmetries of the coordination structure.

\begin{figure}[h]
\centering\subfigure[]{\includegraphics[clip,height=4cm]{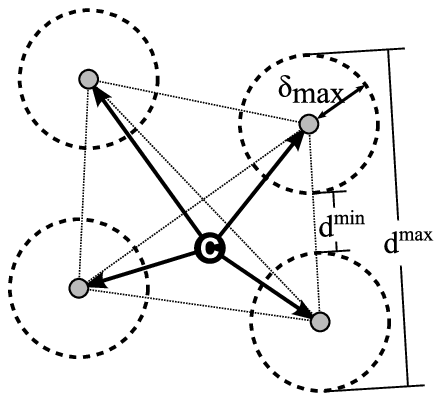}\label{fig:nda_pattern}}

\subfigure[]{\includegraphics[clip,height=4.5cm]{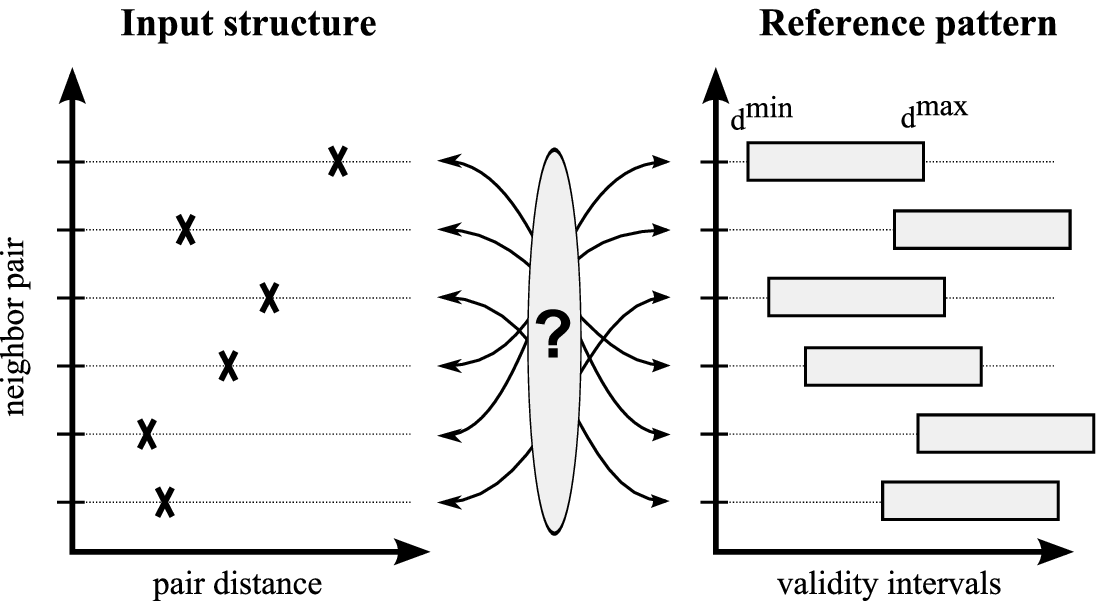}\label{fig:nda_mapping}}\hfill{}\subfigure[]{\includegraphics[clip,height=4.5cm]{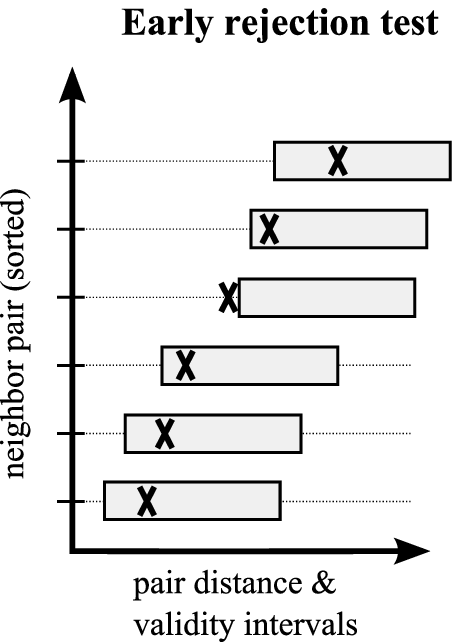}\label{fig:nda_interval_sorting}}

\caption{(a) Schematic picture of a low-symmetry coordination structure. Dashed
circles indicate the maximum distance a neighbor may deviate from
its equilibrium position. This yields six min-max constraints on the
mutual distances between the four neighbors in the example. (b) For
a positive match, a permutation of the neighbors must exist such that
the actual distances fall into the intervals of the reference pattern.
(c) By sorting the neighbor distances and the reference intervals,
a quick rejection test can be performed without knowledge of the actual
neighbor-to-reference mapping.}
\end{figure}

How is the permutation mapping $\sigma$ determined? For this we define
a new type of signature that is based on the linear distance $d_{ij}=\left|\mathbf{r}_{i}-\mathbf{r}_{j}\right|$
between two neighbors $i$ and $j$ of the central atom, which is
invariant under rotation. Hence, we give this approach the name \emph{neighbor
distance analysis} (NDA). We need to consider that particle positions
may be displaced due to thermal vibrations or elastic distortions
of the crystal. Let the maximum allowed deviation of an atom from
its equilibrium position be given by a user-definable parameter $\delta_{\textrm{max}}$.
Then the test structure $(\mathbf{r}_{1},\ldots,\mathbf{r}_{N})$
matches the reference pattern if at least one mapping $\sigma$ exists
such that the condition 
\begin{equation}
\underbrace{\left|\mathbf{R}_{i}-\mathbf{R}_{j}\right|-\delta_{\textrm{max}}}_{d_{ij}^{\textrm{min}}}\;\leq\;\underbrace{\lambda\left|\mathbf{r}_{\sigma(i)}-\mathbf{r}_{\sigma(j)}\right|}_{d_{ij}}\;\leq\;\underbrace{\left|\mathbf{R}_{i}-\mathbf{R}_{j}\right|+\delta_{\textrm{max}}}_{d_{ij}^{\textrm{max}}}\label{eq:interval_cond}
\end{equation}
is fulfilled for all $N(N-1)/2$ neighbor pairs. That is, all rescaled
distances must lie in the corresponding intervals of the reference
structure. This condition is illustrated in figure~\ref{fig:nda_pattern}.

To find a valid permutation map $\sigma$ that fulfills condition
\ref{eq:interval_cond} (or to confirm the non-existence), up to $N!$
possible permutations of the neighbors must tested (figure~\ref{fig:nda_mapping}).
To avoid a fully exhaustive search, the search space can, however,
be considerably reduced by pruning the combinatorial search tree and
employing a backtracking algorithm \citep{ArtOfComputerProgramming4A}.
As an additional optimization step prior to the full combinatorial
search we perform an early rejection test on the entire coordination
structure by sorting both the list of pair-wise distances, $\left\{ d_{ij}\right\} $,
and the list of distance ranges, $\left\{ \left[d_{ij}^{\mathrm{min}},d_{ij}^{\mathrm{max}}\right]\right\} $,
in ascending order (figure~\ref{fig:nda_interval_sorting}). If any
of the distances falls outside the corresponding admissible range,
no valid permutation map can exist and the test structure does not
match the reference pattern.

The user needs to specify two control parameters for each NDA reference
pattern: The number of nearest neighbors to be taken into account
($N$) and the maximum admissible displacement ($\delta_{\textrm{max}}$).
$N$ must be at least three, should include complete shells, and,
apart from that, be as small as possible for best efficiency.

The maximum admissible displacement $\delta_{\textrm{max}}$ determines
the tolerance of the identification process. In general one wants
to use a large $\delta_{\textrm{max}}$ to make the recognition of
structures robust at high temperatures or in the presence of strong
elastic distortions. On the other hand, an excessively large $\delta_{\textrm{max}}$
parameter may lead to \emph{false positives} when testing against
multiple, only slightly different coordination patterns. 

Note that we proposed the NDA primarily for identifying defective
coordination structures that cannot be handled well with existing
methods. In simple cases (such as perfect fcc, hcp, or bcc lattices),
the conventional techniques such as the CNA are the more economic
choice. One important advantage of the NDA, however, is its capability
to identify a wide range of coordination structures. In contrast to
the bond angle analysis method, for instance, which employs hard-coded
decision rules, the catalog of reference structures recognized by
the NDA can be easily extended. The user simply has to provide a set
of perfect reference coordinates, from which the NDA signature for
an atom can be automatically generated. 

We demonstrate this for a molecular dynamics simulation of a $\Sigma11$
$\left\langle 101\right\rangle \left[113\right]$ symmetric tilt grain
boundary \citep{Koning2003a} in fcc aluminum. This low-energy grain
boundary (GB) is composed of repeating structural units, which can
be well identified with the NDA. Two different coordination structures
occur in a perfect $\Sigma11$ $\left\langle 101\right\rangle \left[113\right]$
GB. Hence, the reference pattern catalog contains two GB-specific
structures in addition to the perfect fcc and hcp signatures. The
latter is needed to identify atoms in fcc instrinsic stacking faults,
which exhibit an hcp-like coordination structure. The maximum displacement
parameter is set to 21\% of the nearest neighbor distance, i.e. $\delta_{\textrm{max}}=0.21R_{1}$.

\begin{figure}
\begin{centering}
\includegraphics[width=0.76\textwidth]{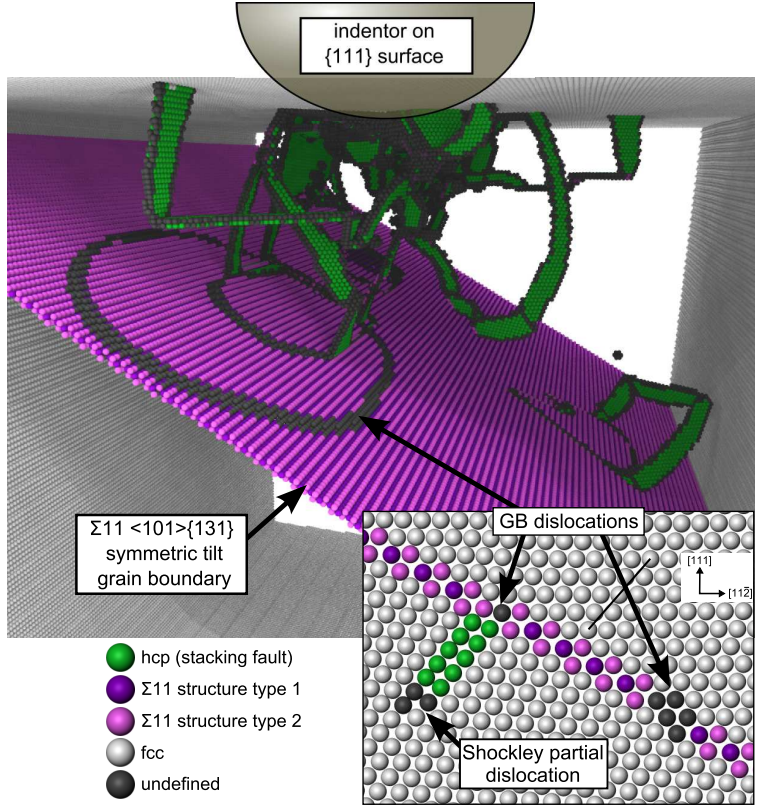}
\par\end{centering}

\caption{Molecular dynamics simulation of a nanoindentation experiment. The
bicrystal contains a $\Sigma11$ $\left\langle 101\right\rangle \left[113\right]$
symmetric tilt boundary that acts as a barrier for lattice dislocations.
The neighbor distance analysis was used to identify undisturbed grain
boundary regions (purple atoms), thereby revealing the cores of secondary
grain boundary dislocations (dark gray atoms), which are the product
of reactions of lattice dislocations with the boundary.}
\label{fig:sigma11gb}
\end{figure}

The simulation setup consists of a bicrystal with a single $\Sigma11$
GB and a nanometer-sized spherical indentor tip. The prismatic dislocation
loops nucleated beneath the indentor interact with the GB (absorption,
transmission, and re-emission). Figure \ref{fig:sigma11gb} shows
the NDA analysis results visualized with \noun{OVITO} \citep{StukowskiOvito2009}.
In the large picture, fcc atoms have been removed to reveal all crystal
defects. One can observe GB dislocation loops gliding in the $\Sigma11$
boundary. The inset shows a cross-section of the symmetric tilt GB
with two secondary GB dislocations. Grain boundary dislocations are
clearly visible because the characteristic structure of the GB is
disturbed inside their cores (dark gray atoms).

\section{Comparison}

The various structure identification methods discussed in this article
employ different types of descriptors or signatures to identify atomic
coordination structures. In general, the classification of a structure
is not based on the particle coordinates themselves but rather on
a derived descriptor. The size of this signature differs for each
analysis method as shown in Table \ref{tab:comparison}. While the
centrosymmetry parameter is a scalar quantity, the neighbor distance
analysis method takes into account all pair-wise distances between
the $N$ neighbor atoms. In general, the capability of a characterization
method to discriminate between a wide range of structures requires
a signature with a sufficient number of degrees of freedom.

\begin{table}
\begin{centering}
\begin{tabular}{|l|c|c|}
\hline 
 & Dimensionality  & Computational\tabularnewline
 & of signature & cost factor\tabularnewline
\hline 
\hline 
Atomic energy & $1$ & -\tabularnewline
\hline 
Centrosymmetry parameter & $1$ & 1\tabularnewline
\hline 
Common neighbor analysis & $3N$ & 3\tabularnewline
\hline 
Adaptive common neighbor analysis & $3N$ & 4\tabularnewline
\hline 
Bond angle analysis & $8$ & 4\tabularnewline
\hline 
Bond order analysis & $2$ & 100\tabularnewline
\hline 
Voronoi analysis & $4$ & 50\tabularnewline
\hline 
Neighbor distance analysis & $N(N-1)/2$ & 20\tabularnewline
\hline 
\end{tabular}
\par\end{centering}

\caption{Comparison of structure identification methods discussed in this article.
The signature dimensionality is the number of components of the vector
that identifies a particular coordination structure, and which is
used for matching within the given analysis framework. $N$ denotes
the number of near neighbors of a central particle. The cost factors
are expressed relative to the computational cost of calculating the
centrosymmetry parameter and do \emph{not} include time spent on generating
neighbor lists. Note that timings are approximate, depend on the input
dataset, and are based on our own, optimized implementations of the
algorithms, which we provide for reference. }
\label{tab:comparison}
\end{table}

We have implemented all analysis algorithms discussed in this article
within a single computer code framework to facilitate the comparison
between them. The code is made available for download at the website
http://asa.ovito.org/. This may be useful for researchers that wish
to further explore comparisons between the methods or for someone
trying to understand the details of the implementations. To measure
and compare their computational costs we applied all discussed methods
to the Fe-Cu dataset shown in figure~\ref{fig:9R_particle}. For
those analysis algorithms that assign a structural type to each atom,
we have included fcc, hcp, and bcc as possible candidates. Calculating
the centrosymmetry parameter is the least expensive analysis, and
we have taken it as reference for our timings. Accordingly, the computation
time per atom of the other methods (Table~\ref{tab:comparison})
is expressed in terms of multiples of this reference time.

\section{Outlook}

Note that the list of methods discussed here is not exhaustive. Indeed,
there exist many more methods, with new ones still appearing, such
that a truly exhaustive study is beyond our finite capabilities. The
aim of the present work was to focus on often-cited techniques that
are routinely used in current simulation studies.

Filtering methods such as the CNA or the CSP are efficient and convenient
techniques that serve well in the visualization and interpretation
of datasets obtained from molecular dynamics simulations of simple
systems with fcc, hcp, or bcc structure. All available structure identification
methods have several limitations in common though, which should be
addressed by future work. By taking into account only near neighbors
of a central atom, the described methods are effectively limited to
simple lattices with a monatomic basis, where the characterization
of the short-range structure around individual atoms is sufficient.
For identification of complex lattices with multiple atoms per primitive
cell (such as 9R) one needs to take into account the medium-range
order of atoms. The same applies to the automated identification of
structured crystal defects such as coherent grain boundaries with
large $\Sigma$, whose characteristic structural units may comprise
many atoms with each having a different local environment.

Furthermore, the sensitivity of structure recognition methods to perturbations
of the particle positions is a problem that hampers the analysis of
systems at high temperature or under large deformation. While the
effect of thermal displacements can, in many cases, be effectively
mitigated by the use of time-averaged particle positions or by quenching
the system using a steepest-descent technique, non-uniform lattice
strains can easily interfere with the identification of coordination
structures. The reason is that most structure signatures used to identify
atomic arrangements are invariant only under rotation but not under
arbitrary affine deformations.

So far, the structure characterization techniqes described in this
article are primarily used to filter simulation datasets to reveal
crystal defects for visualization purposes. In addition, they are
employed to estimate crystal defect densities in MD simulations (e.g.
fcc stackings faults and twin boundaries \citep{StukowskiEtAlTwinPaper2010},
or dislocations \citep{Sansoz2011}). More recently, however, they
have become integral parts of several sophisticated analysis and simulation
methods. Examples for such applications are the characterization of
dislocation lines via an automated Burgers circuit analysis \citep{Stukowski2010},
the mapping of a crystal to a stress-free configuration to separate
elastic from plastic deformation \citep{StukowskiStrainAnalysis},
and the automated construction of a catalog of structural motives
for the efficient discovery of transition events in self-learning
kinetic Monte Carlo simulations \citep{ElMellouhi2008}. Such applications
usually require more than simple classification of local atomic arrangements.
For instance, to determine the local crystallographic directions in
a crystal \citep{Hartley2005}, it is necessary to map all neighbors
of the central particle to the reference pattern in a one-to-one fashion
(as it is already performed by the neighbor distance analysis described
in section~\ref{sub:Neighbor-distance-analysis}), and to determine
the list of all equivalent neighbor permutations, which correspond
to the elements of the point symmetry group of the structure at hand.

\section*{Acknowledgments}

The author thanks Paul Erhart and Tomas Oppelstrup for helpful discussions.
This work was performed under the auspices of the U.S. Department
of Energy by Lawrence Livermore National Laboratory under Contract
DE-AC52-07NA27344.

\section*{\bibliographystyle{unsrt}
\bibliography{Literature}
}
\end{document}